\documentclass[aps,preprint,amsmath,amssymb,showpacs]{revtex4}
\usepackage{graphics}
\usepackage{graphicx}
\usepackage{bm}
\begin{document}
\title{Pancharatnam Phase and Photon Polarization Optics}
\author{S.C. Tiwari}
\affiliation{1 Kusum Kutir, Mahamanapuri, Varanasi 221005, India}
\email{vns_sctiwari@yahoo.com}
\begin{abstract}
Parallel transport of a vector around a closed curve on the surface of a sphere 
leads to a direction holonomy which can be related with a geometric phase that is 
equal to the solid angle subtended by the closed curve. Since Pancharatnam phase 
is half of the solid angle subtended by the polarization cycle on the Poincare 
sphere, quantum parallel transport law takes recourse o spin-half wave function 
to obtain this result. A critique is offered on this factor of half anomaly in 
the geometric phase, and a natural resolution using Riemann sphere polarization 
representation is suggested. It is argued that spin angular momentum of photon 
is fundamental in polarization optics, and new insights are gained based on the 
hypothesis that two helicity states correspond to two distinct species of photon. 
This approach leads to the concept of a physical Poincare sphere: nonlinearity and 
jumps in the Pancharatnam phase find a simple physical explanation while novel 
features pertaining to the discrete and pulsating sphere are predicted. Paired 
photon spin zero structure of unpolarized light is also discussed. An outline 
of possible experimental tests is presented.
\end{abstract}
\maketitle

\section{INTRODUCTION}
Recent advances in the quantum information science have in a subtle way led to a 
paradigm shift on the quantum versus classical debate and controversies; now typical 
counter-intuitive aspects of quantum theory are viewed as resources for novel 
applications. It is recognized that single photon states, quantum vacuum and 
entangled photon pairs constitute one of the most important resources in this 
endeavor, however polarization quantum optics still depends on the classical 
notions developed prior to the emergence of the electromagnetic theory of light. 
One can marvel on the ingenuity of Stokes \cite{1} that the vector nature of light 
could be operationally defined solely based on the intensity (a scalar quantity) 
measurements. In fact, even today introducing polarization states of light without 
any recourse to the electromagnetic fields gives useful insights \cite{2}. Stokes 
vectors have found vast applications in the light scattering experiments, and 
suitable density matrix representation of Stokes four-vector is used to analyze 
polarization-sensitive cross sections in particle physics \cite{3}. In quantum optics 
the Stokes parameters are defined as the expectation or mean values of the Stokes 
operators constructed from the canonical field operators \cite{4}. Note that 
polarization is simply an index attached to the field operators corresponding 
to a quantized simple harmonic oscillator. In the basis of circular polarization, 
the spin angular momentum operator for a plane wave is diagonal in the number 
states, and the difference between the number of right circularly polarized (RCP) 
and left circularly polarized (LCP) photons determines the spin angular momentum, 
see Ch. 10 of \cite{4}. Though mechanical interpretation of classical electromagnetic 
fields relates polarization with angular momentum, and Poynting's suggestion \cite{5} 
found unequivocal support in the Beth experiment \cite{6}, there exist delicate 
questions related to gauge invariance and manifest Lorentz covariance. The problem 
of time-like and longitudinal field excitation is circumvented in quantum 
optics working in the radiation gauge. However, the separation of angular 
momentum of radiation into orbital and spin parts is not unambiguous. 
In most cases plane wave approximation and polarization index associated 
with the field operators imply that it is only the spin of quantized field 
that is of significance. In recent past studies on optical vortices and 
singular laser light beams have brought into focus the significance of the 
orbital angular momentum of light, see the review \cite{7} and references cited therein.
	
A comprehensive critique on the concept of photon \cite{8} concludes that in spite of 
the enormous advances in quantum optics the physical reality of photon remains undecidable.
Should we go back to the basics and look afresh on the physical interpretation of 
the Maxwell field equation? In \cite{8} plausible arguments are put forward to suggest 
that the Maxwell action represents the rotational dynamics of photon fluid. 
I believe that spin angular momentum and polarization hold the secrets of photon 
structure \cite{9}, and going beyond the co-existence of quantum and classical 
descriptions of light there is a need to develop a unitary picture for optical phenomena. 
Recently there has been a renewed intense activity to develop insightful approaches 
to the polarization of light see \cite{10,11,12} and the cited literature. Karassiov 
incorporates SU(2) polarization symmetry at the quantization level \cite{10}. 
Lehner et al \cite{11} investigate unpolarized light employing rotational and 
retardation invariance, and offer a critique to the new class of unpolarized light
discussed in \cite{10}. Luis \cite{12} in a series of papers has developed a formalism
based on the probability distribution on the Poincare sphere to characterize the degree 
of polarization of light.

The aim of the present paper is to revisit Pancharatnam phase which is an 
`all polarization effect', and gain new insights into the nature of light. It is 
well known that Pancharatnam's work \cite{13} was rediscovered after the Berry phase , 
\cite{14} and since then quantum as well as classical explanations to the geometric 
phases in optics have been discussed in the literature. Experimentally numerous 
studies have confirmed the existence of this effect in polarization optics, therefore 
we proceed in the other way and ask: what could be learnt on the properties of light 
from Pancharatnam phase? The main results of our study are as follows. First it is pointed 
out in the next section that there is a discrepancy of a factor of half in the phase 
obtained in modern approaches, and invoking spin-half for two-level polarization system 
is an artifact. In Sec. III a satisfactory formal resolution is established based on 
the Riemann sphere representation of polarization \cite{15}. Postulating that LCP and RCP 
photons are distinct species, spinor wave function and geometrical mapping of polarization
states are analyzed to seek a probability distribution of the number of photons 
corresponding to the states on the Poincare sphere in Sect. IV. Note that elementary
particle physics oriented photon and anti-photon idea was earlier discussed by 
Good \cite{16}, and in the multivector language advocated by Hestenes \cite{17} 
plane electromagnetic wave solutions having positive (negative) frequency correspond 
to RCP (LCP) light In contrast to them, in our extended space-time model of photon, 
the internal structure of photon determines its spin angular momentum and intrinsic
frequency  (energy) \cite{9,18}. The significance of the results obtained in the 
context of nonlinearity and singularity of Pancharatnam phase \cite{19}, 
unpolarized light and spin angular momentum transfer as a physical mechanism 
for geometric phase \cite{20} is discussed in Sect. V.

\section{POINCARE SPHERE AND PANCHARATNAM PHASE}
\subsection{Pancharatnam's original approach}
Pancharatnam's motivation in \cite{13} is to understand the physics of crystal 
optics: interference of polarized light beams, geometrical approach to the polarization 
phenomenon, and spherical trigonometry are the principal ingredients in his approach. It 
is remarkable that the operational definition based on the measurement of intensity of 
light after its passage through the polarizer and analyzer settings pioneered in 
19$^{th}$ century is followed by Pancharatnam. The following proposition 
is proved by him: the interference between mutually coherent light beams of intensities 
$I_1$ and $I_2$ in the polarization states $P_1$ and 
$P_2$ respectively is given by the expression (that defines the phase 
difference $\delta$)
\begin{equation}
I=I_1 + I_2 + 2\sqrt{I_1I_2}\cos P_1 P_2 \cos \delta 
\end{equation}
Here polarization states are described on the Poincare sphere, and 
$P_1$, $P_2$ is the angular separation between the points 
$P_1$ and $P_2$ on the surface of the sphere. Next, the most 
important geometrical result is obtained: a phase of pure geometric origin, $\Gamma$ depends
on the solid angle $\Omega$ subtended by the triangle $P_1 P_2$ P on the
Poincare sphere and is given by
\begin{equation}
\Gamma = \pm \frac{1}{2} \Omega (P_1 P_2 P)
\end{equation}

In the modern derivations the parallel transport of a vector on sphere naturally leads 
to the value of $\Gamma$ equal to $\Omega$, and it is this anomaly of a factor of half that we wish 
to elaborate since it is not sufficiently recognized in the literature. For this purpose 
it is important to realize how a vectorial property of polarization is mapped on to the 
surface of a sphere defined in terms of the scalar quantities: the intensity I, and the 
Stokes parameters (M, C, S). Adopting the four-vector notation (I, M, C, S) can be written 
as $S_\mu$ , $\mu$ = 0, 1, 2, 3. For a perfectly polarized light
\begin{equation}
S_{\mu} S^{\mu} = 0 \qquad \qquad {\rm or} \qquad \qquad I^2 = M^2 + C^2 +S^2 
\end{equation}
Eq. (3) defines the Poincare sphere, and the cartesian coordinates of a point on the 
sphere are given by the Stokes vector {\bf S} (or $S_i$, $i$ = 1,2,3). To make 
transparent the relationship with polarization, consider a general polarization state 
that is described by the orientation of, say major axis of the ellipse, and the ratio 
of semi-minor axis to semi-major axis ($b/a$). Let the propagation direction of light be 
fixed, and it is assumed along z-axis, and the orientation is specified by the angle $\lambda$ 
made by the major axis with the x-axis. Define the ellipticity by an angle $\xi$ such that 
tan$\xi$= b/a. The angles $\lambda$ and $\xi$ called azimuth and ellipticity uniquely represent the 
polarization state; here $0 \leq \lambda \leq \pi$ and  $-\pi/4 \leq \xi \leq \pi/4$.

On the surface of the unit sphere shown in Fig.1, standard geometric definitions 
are introduced: by virtue of Eq.(3) only two coordinates ($\theta,\phi$ ) of spherical polar 
coordinate system are sufficient to specify any point P on the sphere. Polar angle $\theta$ 
is the great circle arc length ZP, and the azimuthal angle $\phi$ is the spherical angle XZP. 
The diametrically opposite points Z and Z' represent the poles of the great circle 
termed the equator. The great circle arc ZP and the spherical angle AZP define the 
latitude [($\pi$/2)-ZP] and longitude respectively. The meridian ZPBZ' intersects the 
equator at B: all points on meridian have the same longitude. The points on the small 
circle KPL with poles (Z, Z') have the same latitude, and it is called the parallel 
of latitude.

\begin{center}
\begin{figure}
\includegraphics[width=4.0in,height=2.5in]{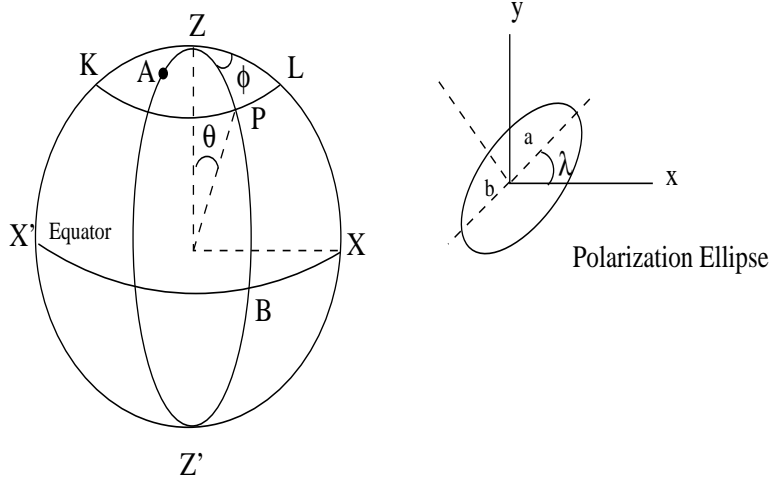}
\caption{Description of spherical surface. Poles $Z$ and $Z'$ represent RCP and LCP
light on the Poincare sphere, and points on the equator $X'BX$ represent linear 
polarization. Polarization ellipse is shown as inset.}
\end{figure}
\end{center}
On the Poincare sphere, latitude and longitude have the values 2$\xi$ and 2$\lambda$ of the 
polarization ellipse shown as the inset in Fig.1. Poles Z($Z^\prime$) represent RCP (LCP) light, 
and points on the equator represent linearly polarized light. Elliptical polarization 
corresponds to rest of the points on the surface, and orthogonal states lie on the 
antipodal points on the sphere. Now it is not necessary to go into the details of 
the derivation of expression (2) for $\Gamma$ which can be found in \cite{13}, the crucial point 
is that a polarization rotation in real space corresponds to a rotation twice of that 
value on the Poincare sphere and Pancharatnam's approach takes into account this 
naturally while considering decomposition of a coherent polarized light beam into 
two parts having arbitrary polarizations. This results into a geometric phase acquired 
by a light beam traversing a cycle along the geodesic path on the Poincare sphere equal 
to half of the solid angle subtended by the closed cycle.

\subsection{Discrepancy of a factor of half}
In a lucid paper \cite{21} Berry brought out the significance of Pancharatnam 
phase in the light of quantum mechanical adiabatic phase discovered by him in 1984 \cite{14}
establishing a close relation of light polarization description on the Poincare sphere 
with the spinor evolution for the Hamiltonian
\begin{equation}
H({\bf r}) = {\bf r}\cdot{\bf\sigma}
\end{equation}
	
Here ${\bf r}$ is a unit vector parametrized by polar angles ($\theta$,$\phi$) and $\sigma$ 
is the Pauli spin matrix. 
In the concluding section of \cite{21} spin-half representation for getting $\Gamma$ 
for photon which 
has spin one is sought to be justified with the argument that photons have only two states 
of helicity. On the other hand, the phase acquired on the closed path on the surface of 
the sphere equal to the solid angle subtended by the path can be identified with the spin 
redirection phase on the sphere of directions of the propagation vector, k. The question 
arises: could SU(2) symmetry associated with spinor be used for spin one photon? In \cite{22} 
we made a brief comment on this question and rather naively suggested that since phase 
corresponds to electric field vector one has to take the square root of phase factor 
obtained on the sphere. In the next section an attempt is made to give a sound basis 
for this suggestion. But first we elaborate the problem.

It is known that non-Euclidean geometry of the sphere results into the change of 
direction of a vector under parallel transport: it can be proved calculating the 
Christoffel symbol and using geodesic equation that a vector parallel transported 
around a curve gets rotated through an angle equal to the solid angle subtended by 
the area enclosed by the curve. In another description, one can use the Fermi-Walker 
parallel transport for curved geometry \cite{23} and obtain the same result for the sphere. 
In the context of geometric phase in optics, Chiao and Wu \cite{24} proposed k space as the 
parameter space and predicted rotation of polarization for a light beam propagating 
through a helically wound optical fiber, and immediately this was experimentally 
demonstrated by Tomita and Chiao \cite{25}. In the light of earlier anticipations due to 
Rytov and Vladimirskii, we term this spin redirection phase as RVCW phase \cite{26}. 
Haldane explained this phase in terms of the geometry of fiber relating it with 
the work of Ross \cite{27}. However, Segert \cite{23} and others have explained the RVCW phase 
purely in geometric terms i.e. tangent bundles and parallel transport laws \cite{28}. In 
\cite{29} Berry discusses parallel transport of a vector on the surface of a sphere, and 
to translate the direction holonomy to phase holonomy he defines a complex unit vector 
and chooses local basis vectors along the parallel of latitude and meridian of longitude. 
He obtains the expected result, namely the phase holonomy is equal to the solid angle 
subtended by the closed curve on the sphere. Obviously the RVCW phase finds natural 
geometric explanation in all these modern approaches \cite{21, 23, 28, 29}.

To derive Pancharatnam phase Berry replaces the complex unit vector by a quantum 
state, and assuming the spinor quantum state for photon arrives at a factor of half. 
Jordan \cite{30} similarly states that, `since the spin eigenvalue is 1/2, the phase
difference is $\Omega$/2'. Bhandari \cite{31} asks the question: Is there any paradox in a spin 1 
particle like a photon behaving as a spin 1/2 particle? He answers, `there is in fact none'. 
The reasoning behind this assertion is two-fold: spin-half representation holds 
for any two-level system, and Jones calculus has been effectively in use for light 
waves in polarization optics. Note that complex representation of the components of 
electric field vector in Jones calculus does not in any way imply its interpretation 
as a wavefunction of photon, and as shown by Jiao et al \cite{32} without 
using any quantum rendition Pancharatnam phase can be derived using Jones calculus. At a basic level, 
there is an intricate problem in defining a wavefunction for photon \cite{8}. As for the 
two-level system, spinor form is merely an analogy; there is nothing quantum mechanical 
in it; the fact partly admitted by Bhandari \cite{31} when he says that quantum statistics 
is different for spin-half particle. The inevitable conclusion is that phase holonomy 
on the sphere (in k space) unambiguously corresponds to the RVCW phase, and Pancharatnam 
phase derived using spinor wavefunction on the Poincare sphere is an artifact.

\section{RIEMANN SPHERE}
Penrose \cite{15} has remarked that the fundamental role of Riemann sphere is not well 
recognized for any two level quantum system. In the case of photon spin the description 
is abstract, but it is worth considering to understand Pancharatnam phase. Let us note 
some mathematical properties of the sphere \cite{33} which have importance for the present 
discussion. The sphere is defined on the real space i.e. field of real numbers 
$R^3$ and it is orientable. Orientability could be proved considering 
the parallel transport of a frame where frame is a set of linearly independent 
vectors ($u, v$). The atlas defined on the sphere assigns different set of coordinates 
on the two hemispheres. Physically for the Poincare sphere, $R^3$ is 
described by (M, C, S), and orientability is responsible for the sign of the Pancharatnam 
phase in Eq. (2). There cannot exist a continuous, regular vector field on the sphere. 
In a simple illustration this is reflected in the stereographic projection where infinity 
is mapped on to the pole of the sphere. The sphere is not a group manifold, and the group 
manifold of SU(2) is 3-sphere defined on $R^4$. Further the group of rotations 
SO(3) has a relationship with SU(2): the mapping SU(2) $\to$ SO(3) is a two-to-one homomorphism.  
Aitchison \cite{34} gives a nice discussion on the monopole problem and absence of 
singularity-free vector potential in terms of the property of the sphere that continuous 
regular vector field cannot be defined on it. He also shows that the Berry phase-monopole 
relationship can be made more clear using the fact that 3-sphere is a group manifold of 
SU(2). Since SU(2) is double-covering of the rotation group in $R^3$ i.e. SO(3), 
the spin-half particle can be described by SU(2); Berry phase for the neutron using spinor 
wavefunction on the sphere would be unambiguous. 

We propose that parallel transport of a vector on the Riemann sphere representation of 
light polarization leads to the Pancharatnam phase. Note that this phase can be obtained 
from the geometry of the Poincare sphere (without making use of spinor wavefunction and 
quantum parallel transport); see \cite{32} besides Pancharatnam's derivation \cite{13}. 
In \cite{33} it 
is proved that the complex sphere is homeomorphic to the tangent bundle of the (real) 
sphere: could one relate Riemann sphere with the Poincare sphere in a similar fashion? 
We leave aside this question here, and proceed to describe Riemann sphere construction 
following \cite{15}.

Let us consider two complex quantities z and w, and define their ratio
\begin{equation}
q = \frac{z}{w}
\end{equation}

In the Argand plane, a complex number is geometrically represented as a point with real 
and imaginary parts as coordinates along the two orthogonal axes (R,I). Let us construct 
a unit sphere whose centre is taken to be the origin of the Argand plane lying horizontally 
as shown in Fig.2. The points (1, 0, (0, i), (-1, 0) and (0, -i) lie on the equator of the 
sphere. Assuming that one of the poles represents infinity, the projections from this point 
to the Argand plane map all of the numbers q (including the one where w = 0) to the surface 
of the sphere uniquely; i.e. the stereographic projection.

\begin{center}
\begin{figure}
\includegraphics[width=3.5in,height=2.5in]{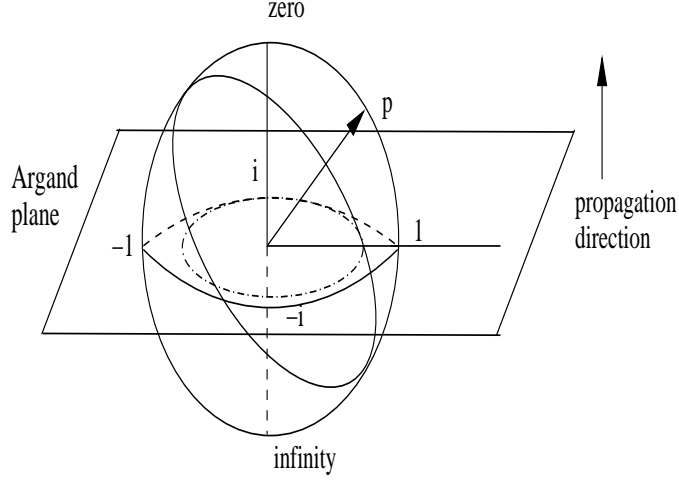}
\caption{Reimann sphere construction-mapping points on Argand plane on to the surface of
sphere. Inset shows propagation direction of light. Dotted ellipse is a projection of the
circle perpendicular to the line connecting centre to the point $p$.}
\end{figure}
\end{center}
To represent polarized light on the Riemann sphere, let us write an arbitrary 
polarized state as a linear combination of RCP and LCP light
\begin{equation}
P = RCP + q LCP
\end{equation}

Assuming that the poles represent RCP and LCP light, the complex number $q$ on the 
Riemann sphere represents an arbitrary polarization state specified by its square root
\begin{equation}
p = \sqrt{q}
\end{equation}

As shown in the inset of Fig. 2 direction of propagation of light is vertically upwards, 
and the polarization ellipse is obtained as follows. The plane normal to the line joining 
the centre to a point on the sphere intersects the sphere in a circle. Projection of this 
circle on to the horizontal plane gives the ellipse which represents the elliptically 
polarized state at that point.

It is now straightforward to obtain Pancharatnam phase: square root of the parallel 
transport phase holonomy on the Riemann sphere immediately gives the correct value 
given by the Eq.(2).To end this section we quote an interesting observation made by Penrose in a footnote 
(p.272) \cite{15}: ``The square root has to do with the fact that the photon is a massless 
particle of spin one, i.e. twice the fundamental unit $\hbar$/2. For a graviton -- the yet 
undetected massless quantum of gravity -- the spin would be two, i.e. four times the 
fundamental unit, and we should need to take the fourth root of q in the 
above description''.

\section{PHOTON SPIN AND LIGHT POLARIZATION}
The term `photon spin' is freely used by Penrose, however in the light of 
conceptual problems and elusive physical realization of photon \cite{2,4,8} we have used 
light polarization in the preceding section to describe the Riemann sphere. It is known 
that to picture linearly polarized photon -- a single photon, one has to rely on the 
counter-intuitiveness of quantum mechanics. Feynman in his characteristic style discussed 
this issue \cite{35} and quantum probability amplitude comes in handy for this purpose. 
That it does not make sense to imagine a fraction of one photon means interpreting the thought 
experiment of the passage of a single photon through polarizer/analyzer differently, e.g. 
he says, `Quantum mechanics tells us it is all there 3/4 of the time'. We argue in this 
section that spin angular momentum of $\hbar$ per photon (demonstrated in numerous
experiments) in the light beams ought to be accorded a fundamental role in the physical model of a 
single photon. Before that we draw attention to the geometric phase unwittingly found 
by Feynman. In the discussion on what he calls a `curious point', an interesting result 
is obtained: RCP (LCP) photon though remains RCP(LCP) photon if viewed from any arbitrary 
rotated frame, the photon acquires a phase factor that distinguishes the frames. Note that 
this observation predates Byrne's analysis \cite{36} that we highlighted in \cite{20}.

We adopt a heuristic approach based on the hypothesis: spin of photon is a 
characteristic property of its internal structure, and spin+ $\hbar$ (or RCP) photon is a 
distinct object than spin -$\hbar$  (or LCP) photon. As a consequence there does not exist a 
linearly or elliptically polarized photon, and RCP photon cannot change to LCP photon 
in a passive optical process. This hypothesis does not depend on any specific model of 
the internal structure of photon and therefore inconclusive state of the photon model 
discussed in \cite{9} is unimportant. In the present paper the polarization optics is 
discussed based on the ensemble of the constituent quantum objects: RCP and LCP photons. 
Intensity of light and polarization state description are the main ingredients for 
this polarization optics. Note that we are not using second quantized formulation of 
optics here. For simplicity we consider light propagation along a fixed direction, 
say z-axis. Total average number of photons N determines the light intensity, I.
A reasonable assumption is that I is proportional to N. Let N$_{\mathrm{r}}$ 
and $N_l$ be the number of RCP and LCP photons, then the problem is 
to characterize the state of light. In a stream of photons possessing on the average equal number of 
RCP and LCP photons, the net spin of the light beam is zero. However, the difference 
in the distribution of photons will give rise to the different states of what we call 
linearly polarized light in this case. The intensity being a scalar is a constant 
quantity for a fixed N therefore to account  for this difference `phase' seems a 
natural choice. To understand this, let us go back to the elementary description of 
a simple harmonic oscillator. To describe its state, say displacement, one can just 
record its instantaneous position at different points of time or alternatively describe 
the motion in terms of a constant amplitude and a phase variable. Intensity is analogous 
to the constant amplitude of the oscillator, and one needs another variable, i.e. phase 
to describe the photon stream. Thus probability distribution of N$_{\mathrm{r}}$ and  
N$_{\mathrm{l}}$,and phase would represent the light in an arbitrary polarized state.

Based on our hypothesis we reanalyze the geometrical description of polarization, 
i.e. the Poincare sphere. The RCP has $N = N_r$, and the LCP light 
$N = N_l$, therefore the poles have fundamental significance representing these states. We restrict 
the domain of the surface of the sphere in two hemispherical regions $S_+(0 \leq \theta < \pi/2 + \epsilon)$ 
and $S_-(\pi/2 - \epsilon  < \theta \leq \pi)$, and the overlap equatorial region has 
non-trivial topology. To characterize probability distribution we reinterpret 
the spinor functions
\begin{eqnarray}
\psi_+ &=&
\left[\begin{array}{cc}
\cos{\theta/2} & e^{i\phi/2} \\
\sin{\theta/2} & e^{-\phi/2}
\end{array} \right]    \\
\psi_- &=&
\left[\begin{array}{cc}
\sin{\theta/2} & e^{i\phi/2} \\
-\cos{\theta/2} & e^{-\phi/2}
\end{array} \right] 
\end{eqnarray}

Multiplying  both sides of Eqs. (8) and (9) by $\sqrt{N}$ we have
\begin{eqnarray}
P_+ &=& \sqrt{N} \psi_+ = \sqrt{N_r} e^{i\phi/2} + \sqrt{N_l} e^{-i\phi/2} \\
P_- &=& \sqrt{N} \psi_- = \sqrt{N_r} e^{i\phi/2} - \sqrt{N_l} e^{-i\phi/2}
\end{eqnarray}
where on $S_+$ and  $S_-$ respectively $N_r$ and $N_l$ are given by
\begin{eqnarray}
\sqrt{N_r} &=& \sqrt{N} \cos{\theta/2}, \qquad \qquad \sqrt{N_l} = \sqrt{N} \sin{\theta/2} \\
\sqrt{N_r} &=& \sqrt{N} \sin{\theta/2}, \qquad \qquad \sqrt{N_l} = \sqrt{N} \cos{\theta/2} 
\end{eqnarray}

Conventionally diametrically opposite points on the Poincare sphere represent 
orthogonal (elliptical) polarization states; in the new interpretation these correspond 
to the light consisting of the mean number of RCP (LCP) photons interchanged with LCP(RCP) 
photons. The equator has a special significance since near the transition overlap region 
the difference between the numbers $N_r$ and $N_l$ tends to 1, and the cross-over 
implies a jump in handedness. For, what one usually refers to as classical light beams, 
besides the averaged field quantities and corresponding Stokes parameters, one can go a 
step further and introduce second order correlations of the fields. The fluctuations in 
fields (and intensities), both spatial and temporal, are inherent in the light 
phenomena; field correlations at space-time points in the coherence theory give 
information regarding such fluctuations. The 2x2 coherence matrix defined for vector waves 
embodies the polarization property of light in this second order coherence theory \cite{4}.
In quantum optics, quantum field operators have to be used, however one may simplify 
the discussion using photon number fluctuations. Evidently such effects would be more 
pronounced near the equatorial region of the Poincare sphere. In the next section some 
of the physical consequences of the present approach are discussed.

\section{NEW INSIGHTS AND PROPOSED TESTS}
The approach proposed in the preceding section is based on the quantum nature of 
light whose constituents are two species of photon distinguished by their uniquely 
defined spin (or handed-ness). Note that we are not using the philosophy of quantum 
mechanics (a la single particle interpretation of the wavefunction), and do not adopt 
second quantized field theory. The photon polarization optics (PPO), being developed, 
is inspired by the geometrical considerations and the extended space-time model of 
photon. Known polarization properties of light are incorporated by construction in 
the PPO, however new insights are also gained that are presented in the following:

{\bf(i)	Unpolarized light:} The definition of unpolarized light is not 
simple in spite of the fact that natural light is unpolarized. The present work suggests 
a definition: uniformly weighted paired states ($P_+$, $P_-$) 
spanning the whole surface of the Poincare sphere characterize the unpolarized light. 
Recall that in any paired state ($P_+$, $P_-$) the number of 
RCP (LCP) photons in $P_+$ is equal to the number of LCP (RCP) photons 
in $P_-$. The pairing gives rise to net spin zero, and the resulting 
distribution of paired photons represents unpolarized light. In \cite{9} it was argued 
that a random ensemble of quantum entangled photons could be viewed as a new state 
of light. Disregarding quantum mechanical interpretation, the unpolarized light 
defined here essentially corresponds to this `new state of light' -- let us call it 
paired photons unpolarized light (PUL).

Luis defines unpolarized light \cite{12} in terms of the Q function having the value
\begin{equation}
Q(\theta,\phi) = \frac{1}{4\pi}
\end{equation}

This corresponds to a uniform distribution for fields. In his work SU(2) coherent states 
are used. Formally our definition also entails 4$\pi$, i.e. the solid angle subtended by the 
whole surface of the Poincare sphere, however the physical ideas are different. 
Lehner et al \cite{11} define two types of unpolarized light. Type II polarized light is 
defined by a distribution function of fields that is rotationally invariant and symmetric 
with respect to RCP$\to$  LCP transformation. Type I unpolarized light besides these 
requirements possesses phase retardation invariance. The PUL defined here satisfies 
all the conditions necessary for type I unpolarized light, however, PUL has a 
polarization structure of paired photons. Karassiov \cite{10} speculates on the hidden 
polarization structure of unpolarized light. Though relating PUL with the formal 
P-scalar light proposed by him \cite{10} is not clear, we disagree with Lehner et al 
that polarization structure of unpolarized light is paradoxical. Interestingly 
Simmons and Guttmann \cite{2}, (p. 91), also find an unusual representation for 
completely unpolarized light. Using the property that diametrically opposite 
points represent orthogonal polarization states the Stokes four-vector for 
unpolarized light can be rewritten as
\begin{eqnarray}
\left[\begin{array}{c}
I \\ 0 \\ 0 \\ 0
\end{array} \right] =
\left[\begin{array}{c}
I/2 \\ M \\ C \\ S
\end{array} \right] +
\left[\begin{array}{c}
I/2 \\ -M \\ -C \\ -S
\end{array} \right]
\end{eqnarray}
Thus unpolarized light is an incoherent superposition of any two polarized states.

The preceding discussion should not be construed to imply that photon pairing 
is a basic requirement for unpolarized light. If we relax the plane wave approximation 
and a fixed direction of propagation then randomly oriented spinning photons in the 
beam such that the net average spin is zero, would also represent unpolarized light. 
This kind of unpolarized light cannot be said to have a polarization structure. 
How to distinguish the two kinds of unpolarized light? It can be easily guessed 
that the PUL has the structure of the correlated quantum entangled pairs, therefore 
the difference between the polarization correlation measurements should throw light 
on these states. Mandel and Wolf \cite{4}, p.649, point out that each one of the photon 
pair in quantum entangled state is unpolarized as defined by the coherence matrix. Thus 
it would seem that these questions deserve deeper analysis, and further study.

{\bf(ii) The nature of Pancharatnam phase:}	The Poincare sphere used in the 
work of Pancharatnam retains its mathematical idealization for continuous field variables 
representing light. The quantum nature embodied in the PPO, on the other hand, gives rise 
to the idea of a physical Poincare sphere in which two poles representing RCP (LCP) light 
are indispensable, while the equator relates the topological property of the sphere. 
Pancharatnam phase for a polarization cycle on either hemi-sphere is geometrical, while 
near the equator due to the constraint that fractional photon is unphysical, one would 
expect peculiar behaviour. Schmitzer et al \cite{19} found such an `exotic' nonlinearity in 
the Pancharatnam phase. The cross-over from the excess of N$_{\mathrm{r}}$ to the excess 
of $N_l$ (or $N_l$ to $N_r$) in the polarization 
path reveals in a topological effect 
of phase jumps. Bhandari's observations \cite{19} are proposed to be due to this effect. Thus 
nonlinearity and singularity in Pancharatnam phase could be given a physical origin in PPO.

The physical mechanism responsible for the geometric phases (including the RVCW
phase) was suggested to be the transfer of angular momentum in \cite{20}. We have elaborated 
this suggestion recently \cite{26} and proposed experiments to test these ideas. In the 
context of Pancharatnam phase, the question that we ask is the following. Let us 
consider a beam of light that has specified average values of $N_r$ 
and $N_l$; after the passage through optical elements that change its polarization 
let it return to the state with the same $N_r$ and $N_l$. 
How to distinguish the two states? Of course, Pancharatnam phase represents a criterion to 
distinguish the final state from the initial state. In \cite{20,26} the notion of the uniform level 
of angular momentum has been discussed, and it is argued that this level is changed that manifests as 
a geometric phase. A quantitative estimate is lacking, however in the light of PPO 
a plausible connection with the rearrangement of the distribution of photons could 
be made. Though average $N_r$ and $N_l$ are same in the initial and final 
states, the local distribution and correlations of photons are changed in the 
rearranged final state of the light. Therefore, we expect that physical Poincare 
sphere has a granular structure (or discretized pairs on the surface) and fluctuating 
radius. To probe the spin transfer mechanism and the hidden structure of the sphere, 
a great deal of experimental ingenuity would be required. It is suggested that 
second-order correlations, and higher-order interference effects for low intensity 
light beams undergoing polarization cycles could probe these features.

An intriguing observation is that of frequency shift as an  evolving 
Pancharatnam phase \cite{37}: a rotating half-wave plate seems to induce a frequency shift.
Bretanaker and Le Floch \cite{38} argued that the experimental observation could be 
explained in terms of the energy conservation. They assume that the light beam consists 
of $N\sigma^+$ (RCP) and $N\sigma^-$ (LCP) photons,
and calculate the energy transfer via rotating half-wave plate. Angular momentum exchange (torque 
imparted to the wave plate by the incident beam) leads to the energy transfer, and 
energy conservation law gives the frequency shift. It is known that for a monochromatic 
radiation no passive electronic/optical element can change its frequency, therefore 
the frequency shift by a rotating wave plate is intriguing. An alternative explanation 
consistent  with the hypothesis that internal frequency of RCP and LCP photons cannot 
be changed unless these are exchanged by photons with different frequency, would be to 
interpret the observed intensity oscillation in terms of the time-dependent photon 
numbers $N_r$ and  $N$ in Eqs. (10) and (11). The question whether the 
time-varying phase change is to be interpreted as frequency shift or oscillations 
in the number of photons cannot be decided solely on the intensity measurements. 
On the physical Poincare sphere the time-dependence of $N_r$ and  $N_l$ should 
manifest as a pulsating sphere, therefore to observe this feature measurements on all 
the Stokes parameters (M, C, S) would be required. Another possibility is to seek a 
detection scheme for the output beam intensity that distinguishes frequency effects 
and the effect of photon numbers.

\section{DISCUSSION AND CONCLUSIONS}
In this paper we have proposed an approach to understand the polarization property 
of light based on a physical model of photon, and yet did not consider single photon effects. 
The reason is two-fold: first physical model of photon is at present tentative, and secondly 
the approach outlined here is in a developing state. There has been a great deal of
understanding on the classical light, and quantum optics for large number of photons.
For a single photon there are many conceptual problems \cite{8}, and the role of quantum 
vacuum is very significant. Localized photon and the spin angular momentum in more 
physical terms (i.e. the structure of photon) than a polarization index have not 
found a satisfactory solution in quantized field theory. The vacuum field becomes 
crucial in quantum optics even in the case of a beam splitter, see Sec. 10.9.5 in \cite{4}. 
The significance of Pancharatnam phase to gain insights has been stressed here. 
In view of the recent speculations on what has come to be known as `quantum-vacuum 
geometric phase', see \cite{39} it would seem that in the context of our hypothesis the 
very idea of Pancharatnam phase for a single photon would involve quantum vacuum. 
Shen \cite{39} discusses RVCW phase for RCP and LCP photons, and argues that the phase 
at quantum-vacuum level could be envisaged. Our hypothesis, on the other hand, implies 
that a single photon can be represented only at one of the poles of the physical Poincare 
sphere, and any notion of a polarization cycle for a single photon depends on the existence 
of a hidden source of photons which could be identified with aether or quantum vacuum. 
Obviously instead of special optical media to realize quantum-vacuum geometric phase 
\cite{39}, here it is suggested that single photon polarization cycle, if observed carefully, would 
reveal the quantum-vacuum effect for Pancharatnam phase and similar to the Casimir force, 
quantum vacuum torque would come into the play.
	
In conclusion, we have elaborated a conceptual problem in the derivation of 
Pancharatnam phase using quantum parallel transport that depends on the spin-half 
treatment for photon having spin one, and suggested that the square root of phase 
factor obtained on the Riemann sphere gives correct result. A heuristic approach 
for polarized light is outlined introducing the hypothesis that RCP and LCP photons 
are distinct species, and spin angular momentum originates due to the structure of 
photon. The notion of physical Poincare sphere, new insights on the unpolarized light, 
and the implications of these ideas on the Pancharatnam phase are discussed. 
Qualitative arguments are also presented to test some of the ideas experimentally. 
The role of spin angular momentum to address fundamental questions seems inevitable, 
both for photon, and electron discussed elsewhere \cite{18,40}.

\section{acknowledgements}	
The library facility of the Banaras Hindu University, Varanasi is  acknowledged.

\end{document}